\documentclass[a4paper,12pt]{article}

\usepackage{setspace}
\usepackage{textcomp}
\usepackage{amsmath,amssymb,amsfonts,fullpage}
\usepackage{subfigure}
\usepackage[dvips]{graphicx}
\usepackage{natbib}

\newcommand{\di}{\partial} 
\newcommand{\apgt}{\ {\raise-.5ex\hbox{$\buildrel>\over\sim$}}\ }
\newcommand{\aplt}{\ {\raise-.5ex\hbox{$\buildrel<\over\sim$}}\ }

\def\Hcal{{\cal H}}

\def\Hcal{{\cal H}}

\def\ex{\rm e}

\begin{document}
\doublespacing

\title{\bf{A 1-D evolutionary model for icy satellites, applied to Enceladus}}

\author{Uri Malamud and Dina Prialnik}

\maketitle
\begin{center}
urimalam@post.tau.ac.il ~~~~ dina@planet.tau.ac.il\\*[1cm]
\end{center}

\date

\maketitle
\begin{center}
Department of Geosciences \\Tel Aviv University\\Ramat Aviv, Tel Aviv 69978, Israel\\*[1.5cm]
\end{center}

\newpage

\begin{abstract}
We develop a long-term 1-D evolution model for icy satellites that couples multiple processes: water migration and differentiation, geochemical reactions and silicate phase transitions, compaction by self-gravity, and ablation. The model further considers the following energy sources and sinks: tidal heating, radiogenic heating, geochemical energy released by serpentinization 
or absorbed by mineral dehydration, gravitational energy and insolation, and heat transport by conduction, convection, and advection. We apply the model to Enceladus, by guessing the initial conditions that would render a structure compatible with present-day observations, assuming the initial structure to have been homogeneous. Assuming the satellite has been losing water continually 
along its evolution, we postulate that it was formed as a more massive, more icy and more porous satellite, and gradually transformed into its present day state due to sustained long-term tidal heating. We consider several initial compositions and evolution scenarios and follow the evolution for the age of the Solar System, testing the present day model results against the available observational constraints. Our model shows the present configuration to be differentiated into a pure icy mantle, several tens of km thick, overlying a rocky core, composed of dehydrated rock at the center and hydrated rock in the outer part. For Enceladus, it predicts a higher rock/ice mass ratio than previously assumed and a thinner ice mantle, compatible with recent estimates based on gravity field measurements. Although, obviously, the model cannot be used to explain local phenomena, it sheds light on the internal structure invoked in explanations of localized features and activities.
\end{abstract}

\newpage

\section{Introduction}\label{S:Intro}

Enceladus is one of the most intriguing icy bodies in our solar system, since it is the only satellite, except possibly Europa \citep{RothEtAl-2014}, on which active cryovolcanism on a large scale has been detected. The uniqueness of Enceladus has prompted wide investigation in order to uncover its mysteries. To understand the evolution of Enceladus, and of similar icy satellites, one should couple all relevant processes into one global model. Such a general model can, of course, be applied to any icy satellite in a similar range of size and composition, but the detailed observations of Enceladus may be used to constrain a complicated model. In what follows in this section we briefly outline the most recent observations that can be used to constrain models. We then show that although there is a wide range of processes that may contribute to the evolution of icy satellites in general and Enceladus in particular, theoretical models tend to addresses these processes separately. Our aim is to develop a comprehensive model, capable of coupling many different processes.

\subsection{Observational Constraints}\label{SS:Constraints}

\subsubsection{Bulk Properties}\label{SSS:Structure}

The shape of Enceladus deviates slightly from a spheroid \citep{ThomasEtAl-2007}. It has a mean radius of 252.1$km$, and a bulk density of 1.609 $g \cdot cm^{-3}$ \citep{JacobsonEtAl-2006,RappaportEtAl-2007}. 
The bulk density measurement provides an important clue to its interior composition. Interestingly, the satellite Mimas, for comparison, which is similar in size to Enceladus, has a bulk density of only 1.15 $g 
\cdot cm^{-3}$ \citep{JacobsonEtAl-2006}, indicating either a lower rock/ice mass ratio, or a more porous structure (or both). Recently, \cite{IessEtAl-2014} also measured the moment of inertia of Enceladus to be 0.335, suggesting a differentiated global structure. 

\subsubsection{Plume -- Water Loss}\label{SSS:Plume}

The most intriguing discovery of the Cassini orbiter is the south polar anomaly of Enceladus. The spacecraft detected very young features on the surface of the south pole region, out of which plumes dominated by water gas were emanating \citep{PorcoEtAl-2006,SpencerEtAl-2006,WaiteEtAl-2009}. The plume is believed to originate from four prominent fissures, also known as the "tiger stripes". The water flux estimation was obtained when the Cassini ultraviolet imaging spectrograph (UVIS) observed two stellar occultations and one occultation of the sun by the plume above the south polar terrain (SPT) \citep{TianEtAl-2007,HansenEtAl-2011}. In all three occultations, the water fluxes were determined to be comparable, around 150-200 $kg \cdot s^{-1}$. 

It is important to keep in mind that the similar flux measurements obtained do not necessarily imply water flux invariability \citep{HurfordEtAl-2007}. As it happened, all three occultations occurred when Enceladus was at a similar 
orbital phase in its motion, which may account for the fact that the measured fluxes were similar \citep{HedmanEtAl-2013,NimmoEtAl-2014}. The existing observations of water flux are only correct to first order approximation, and should be regarded with caution. Nevertheless, a viable model should account for the possibility of water loss in an appropriate amount.

\subsubsection{Heat Flux -- Surface Temperature}\label{SSS:HeatFlux}

The surface temperature detected in the SPT was identified early on to be much higher than the expected insolation equilibrium temperature \citep{SpencerEtAl-2006}. On April 14, 2012, the Cassini spacecraft performed 
a close flyover of the South Pole, observing the 'Baggdad Sulcus' fissure in very high resolution using the VIMS instrument. \cite{GoguenEtAl-2013} interpreted the obtained spectrum as thermal emission from a linear 
fissure with a temperature of approximately 197 $\pm$ 20 $K$ and width of about 9$m$. Globally, the equilibrium surface temperature is much lower, around 70$K$. \cite{TravisSchubert-2015} suggest that an insulating layer of very porous 'snow' at that region could naturally push the near surface temperatures to the observed level, due to its significantly lower thermal conductivity.

In order to constrain the total power radiated from the SPT, one must consider the integrated thermal emission of the south pole, after subtraction of the estimated passive emission due to re-radiated absorbed sunlight. 
\cite{SpencerEtAl-2006} have estimated an integrated SPT nonsolar power of 5.8 $\pm$ 1.9 $GW$. Higher resolution observations in 2008, gave a higher heat flow estimate of 15.8 $\pm$ 3.1 $GW$ \citep{HowettEtAl-2011}. A more recent study by \cite{SpencerEtAl-2013} argues that previous estimates did not properly resolve the tiger stripes from their surroundings, and that the radiated power from the SPT is only be about 4.7 $GW$. These results should be taken into account by appropriately determining the boundary conditions of the model.

\subsection{Thermo-physical Processes}\label{SS:Models}

\subsubsection{Tidal Heating}\label{SSS:TidalHeating}

Most evolution models for Enceladus concentrate on tidal heating, due to its substantial long-term contribution. Simple models of tidal heating are generally concerned with the total heat dissipated globally \citep{SquyresEtAl-1983}. If the satellite is in synchronous spin-orbit resonance with negligible obliquity, orbital inclination, as well as physical libration, then the global energy release rate is given by a simple expression. The rate not only depends on the physical and orbital parameters of the satellite, but also on the capacity of the satellite to deform and dissipate mechanical energy, given by the effective dimensionless ratio $k_2/Q$, where $k_2$ is the dynamic Love number, and Q the tidal dissipation factor (see eq. \ref{eq:tidal_global} in section \ref{SS:Tidal}). The parameter $k_2$ is essentially a proportionality coefficient, measuring the tidal response to the external tidal potential. In a simple rigid homogeneous body or a fluid-like body, the value of $k_2$ is trivially determined. The parameter Q is related to the amount of energy dissipated through friction, so it too depends on internal variations \citep{RambauxCastillo-Rogez-2013}. The simple expression for the global energy dissipation rate does not provide information about the exact distribution of heat, which is why more sophisticated calculations were carried out. These are, however, complicated and time-consuming, thus some studies still prefer the simple 1-D global energy dissipation rate \citep{MeyerWisdom-2008,LaineyEtAl-2012}.                                    

Of particular relevance to our simple 1-D model is the 3-D tidal heating model for Enceladus developed by \cite{RobertsNimmo-2008}, following the approach of \cite{TobieEtAl-2005}: they show that tidal dissipation in the core is negligible at any plausible silicate viscosity. The more recent analysis by \cite{Roberts-2015} suggests that under certain conditions it may be small, but not entirely negligible, as suggested in the previous study. Consequently, it may be reasonable to assume that most (but not all) of the tidal energy in Enceladus is uniformly distributed within the ice shell, which is thin compared to the radius of the satellite \citep{Barr-2008}. However, tidal dissipation is also sensitive to viscosity, and since ice viscosity is highly dependent on temperature, the difference in viscosity between the surface and the base of the mantle could be considerable \citep{Barr-2008,HanEtAl-2012}. This, in turn, leads to a difference in tidal dissipation, which has been shown to peak at a temperature of about 250K \citep{HanShowman-2010}. Thus, an inward-increasing rather than a uniform tidal dissipation distribution in the icy mantle is a more 
likely approximation in a 1-D approach. 

\subsubsection{Convection}\label{SSS:Asymmetry}
Convection in the ice shell has been suggested to transport the internally dissipated heat to the surface, leading to the young geological features observed in the SPT. Several highly sophisticated 3-D convection models have been developed \citep{BehounkovaEtAl-2012,ShowmanEtAl-2013,RozelEtAl-2014}, specifically in order to understand Enceladus' hemispheric dichotomy. In our 1-D model however we apply a much simpler, parametrized convection model for the ice shell, following the calculation of \cite{BarrMcKinnon-2007}. \cite{TravisSchubert-2015} use a similar parametrization approach in their long-term simulations; however, they also consider hydrothermal flow (mining energy from the core) having different model assumptions than ours.

\subsubsection{Chemical Alterations}\label{SSS:Chemical}

Serpentinization reactions are exothermic chemical reactions that occur as liquid water reacts with pristine (non-hydrated) silicates. Serpentinization is significant on several accounts \citep{MalamudPrialnik-2013}: first, it alters the rock's density considerably; secondly,  it causes a non-negligible amount of water to be absorbed by the rock, which thus gains mass; and finally, it generates substantial amounts of energy. Thus, in a differentiated body, the density of an aqueously altered core is about 20$\%$ lower compared to the density of a non-hydrated one, and the overall amount of free water in the entire body is lower. The energetic contribution of serpentinization has long been considered important: a simple latent heat argument shows that for every mole of serpentine produced, the energy released is able to melt about 11 moles of water ice. Hence there is the possibility for a reaction runaway, in which heat from serpentinization in a localized region of the body triggers ice melting throughout the body \citep{JewittEtAl-2007}. Serpentinization is also recognized to be consistent with the compositional data on the Enceladus' plume, although it must not necessarily be ongoing \citep{GleinEtAl-2015}.

If serpentinization is allowed to occur, then one must also consider the inverse process of rock dehydration. Once dehydrated, the rock becomes denser, and water is released back into the body, while the rock mass is reduced by the same amount. Thus dehydration has the opposite effect on the global composition and structure. The rate of serpentine dehydration is a function of temperature, 
and normally high temperatures are required (for example, the dehydration rate for the Serpentine polymorph antigorite \citep{SawaiEtAl-2013} is entirely negligible below about 675$K$). Thus dehydration is not expected to occur in objects which are relatively small and cold. In the present model we couple serpentinization and dehydration in the thermo-physical evolution model. It should be noted that more realistic hydrated rocky compositions would not consist solely of Serpentine. For example, a couple of possible compositions with low grain densities and residual porosity were suggested for the dwarf 
planet Ceres \citep{Zolotov-2009,Castillo-Rogez-2011}. Nevertheless, modeling only Serpentine dehydration, while ignoring other mineral hydrates in the mixture of \cite{Castillo-Rogez-2011}, is justified in our case, where emphasis is put on evolution and a greatly simplified mineralogy is adopted. This is because (1) it is the most abundant component in the mixture, (2) its characteristic dehydration temperature falls roughly in between the other constituents in the mixture, (3) its precise temperature-dependent rate is well-constrained empirically.

\subsubsection{Ablation}\label{SSS:Ablation}
Enceladus' plume activity strongly suggests that water, the most abundant volatile \citep{WaiteEtAl-2009}, is slowly depleting. Several studies have discussed the possible mechanisms for generating and sustaining the plume \citep{KiefferEtAl-2006,NimmoEtAl-2007,SchmidtEtAl-2008,GoguenEtAl-2013,PorcoEtAl-2014, YeohEtAl-2015,TuckerEtAl-2015}. It is however difficult to put precise constraints on the rate at which this occurred throughout the satellite's life time (see section \ref{SSS:Plume}). Assuming that the present-day ablation rate has been more or less constant in the last 4.6 $Gyr$, Enceladus 
must have lost around $1/6$ to $1/5$ of its initial mass (for an average water loss rate of between 150 to 200 $kg \cdot s^{-1}$ respectively), and at least $3/4$ of its initial water content. This possibility has been discussed early on by \cite{Kargel-2006}, but only qualitatively. As a working hypothesis and for simplicity, we assume that the same rate of water loss has been maintained throughout evolution. The model developed here is capable of following water migration and outgassing at the surface of the satellite and the resulting ablation by implementing an adaptive-grid numerical technique.

\subsubsection{Porosity and Compaction by Self-Gravity}\label{SSS:Compaction}
Typical icy bodies in the solar system range in size from small comets which are kilometer-sized to large icy objects which are thousands of kilometers in size. Presumably, the larger ones were formed as they accreted the smaller planetesimals. We may take the current estimates of comet densities \citep{WeissmanLowry-2008}, as well as the densities of small to medium sized 
asteroids \citep{BaerEtaAl-2011} to represent those of typical small planetesimals, implying that they  were similarly highly porous. As medium and large icy bodies collect planetesimals to grow in size, self-gravity should be sufficient to compress the pores, so that the post-accretion bulk porosity is not as high as in small comets or asteroids, but porosity should not be altogether eliminated. The more likely effect of self-compaction by gravity is that the porosity profile decreases inward, as demonstrated by \cite{Leliwa-KopystynskiKossacki-2000} for the icy satellite Mimas. 
This also depends on composition, since ice is more susceptible to compaction than rock. \cite{DurhamEtAl-2005} have conducted experiments on compaction of porous water ice at low temperatures, and showed that under conditions relevant to midsized icy bodies, which are made predominantly of water ice, a substantial residual porosity remains. \cite{YasuiArakawa-2009} have argued, based on compaction experiments applied to ice-silicate mixtures, that bodies up to several hundred kilometers in diameter, made up of a mixture of icy and rocky material, should have a very high residual porosity, particularly if temperatures remain below freezing point. 

The rock/ice mass ratio that is commonly suggested for icy satellites, is usually determined from the bulk density, under the assumption that porosity is negligible, which only yields a lower limit. A more accurate determination of the rock/ice mass ratio must take into account the radial distribution of porosity as a function of pressure and temperature. Here we adopt the equation of state described by \cite{MalamudPrialnik-2015}, suitable for porous icy bodies with radii of a few hundred km, which is based on the best available empirical studies of ice and rock compaction. 

A plausible assumption regarding the {\it initial} rock/ice mass ratio, is to adopt the same initial rock/ice mass ratio as Mimas. Being cold and inactive, Mimas is not likely to have experienced internal processing, such as serpentinization or water loss, and thus it is more likely to have kept its original rock/ice mass ratio.

\subsection{Theoretical Models}\label{SS:AllInclusive}
As already mentioned, theoretical studies tend to focus on certain processes separately from others. For example, given the importance of tidal heating,
most studies include complicated 3-D tidal calculations coupled with complex convection models \citep{RobertsNimmo-2008,TobieEtAl-2008,BehounkovaEtAl-2010,BehounkovaEtAl-2012,HanEtAl-2012} and in some cases even coupled with orbital models \citep{MeyerWisdom-2008,RobuchonEtAl-2011,LaineyEtAl-2012,ShojiEtAl-2014}. These are used mainly in order to explain the high heat flux observed and the asymmetric SPT anomaly, yet they ignore differentiation, water ablation, compaction and geochemical reactions, all of which are important long-term processes. Fewer studies do target the early evolution and differentiation phase, focusing on water migration by radiogenic heating \citep{SchubertEtAl-2007,PrialnikMerk-2008,MalamudPrialnik-2013}, and by heat released in geochemical reactions \citep{MalamudPrialnik-2013}. These studies do not include tidal dissipation as an energy source, nor compaction, or -- with the exception of \cite{MalamudPrialnik-2013} -- surface sublimation and outgassing. Some studies have discussed water ablation as a result of the plume, mainly in order to gain insight into the general mechanisms responsible for producing it \citep{KiefferEtAl-2006,NimmoEtAl-2007,SchmidtEtAl-2008,GoguenEtAl-2013},
but not as an integral part of a full thermo-physical long-term evolution model.

A more recent evolution model by \cite{TravisSchubert-2015} couples tidal and radiogenic heating, including convection, but does not consider geochemical reactions, ablation or compaction by self-gravity. \cite{Roberts-2015} iteratively couples two separate codes, one for calculating the tidal dissipation and the other for the thermal evolution, assuming an unconsolidated porous core that retains the same porosity throughout thermal evolution.

In this work we adopt a different approach: our aim is to develop a comprehensive, long-term evolution model, incorporating consistently all the different processes mentioned above, suitable for any icy satellite
around any planet. Such a model will be able to shed light on the internal structure of satellites and their evolution since the time of formation, but --- being 1-D (spherically symmetric) --- it will not be capable of explaining the details of localized phenomena. It will only provide the basis for understanding such localized, temporary phenomena. Enceladus will be used as a test case. 

The rest of the paper is arranged as follows: the details of our model are provided in Section \ref{S:Model}. The model is largely based on an earlier model, presented by \cite{MalamudPrialnik-2015}, with some modifications which are described here in greater detail. A detailed description of a long-term evolution model is given in Section \ref{S:Results}. The results are discussed in Section \ref{S:Discussion}, together with alternative models.

\section{The Model}\label{S:Model}

The model used in this study is based on that developed by \cite{PrialnikMerk-2008} and expanded by \cite{MalamudPrialnik-2013,MalamudPrialnik-2015}. The previous model included the following processes: (1) internal differentiation (2) porosity and compaction (3) serpentinization reactions (4) surface ablation or accretion. In terms of energy sources, it considered nearly the entire energy budget, including: short and long lived radioactive nuclides, latent heat released by serpentinization, surface insolation, gavitational energy released by compaction, and crystallization of amorphous ice. The model treated heat transport by conduction and advection. It also followed the transitions among four phases of water (amorphous ice, crystalline ice, liquid and vapor), 
and two phases of silicates (aqueously altered rock and non-altered rock), accounting for thermal (conductivity, heat capacity) and physical (strength, density) changes in the solid phases. 

Here we consider three new features, which are added to the early model: (1) Serpentine dehydration - leading to energy absorption as well as rock density increase and water release, (2) tidal heating - an internal energy source, and (3) convection - an additional heat transport mechanism.

\subsection{Dehydration}\label{SS:Dehydration}
Dehydration occurs through the transition of aqueously processed rock (hereafter labeled $p$) into unaltered (non-hydrated) rock (hereafter labeled $u$). The derivation of the dehydration governing equations follows the derivation of the serpentinization equations \citep{MalamudPrialnik-2013}, albeit the process is inversed. Water is released back into the system as the consequence of the reaction, there is an increase in the rock density of $\sim$20\% and the heat of the reaction acts as an internal energy sink. Let the number density of $u$ molecules be $n_u$, that of water molecules $n_w$ and of $p$ molecules $n_p$, while their respective molecular weights are $A_u$, $A_w$ and $A_p$, such that $A_p=A_u+2A_w$. The dehydration rate, that is, the rate of change of $n_p$ is given by
\begin{eqnarray}
\label{eq:dehyd1} \dot n_p& = &-n_p R_D \\
\label{eq:dehyd2} \dot n_u& = &-\dot n_p \\
\label{eq:dehyd3} \dot n_w& = &2\dot n_u  
\end{eqnarray}

The dehydration reaction rate $R_{D}$ has the dimension of 1/time. In terms of partial densities, we denote by $\rho_{(p,u,w)}$ the respective masses per unit volume. Since $n$ is related to $\rho$ by $n=\rho/(A*m_u)$, where $m_u$ is the atomic mass unit, eq. (\ref{eq:dehyd1}-\ref{eq:dehyd3}) become

\begin{eqnarray}
\label{eq:dehyd4} \dot \rho_p = -\rho_p R_D \\
\label{eq:dehyd5} \dot \rho_u = -\dot \rho_p \frac{A_u}{A_u+2A_w} \\
\label{eq:dehyd6} \dot \rho_w = \dot \rho_u \frac{2A_w}{A_u}
\end{eqnarray}

For a given $R_D$, eq. \ref{eq:dehyd4} has the form of a standard exponential decay. The dehydration reaction rate $r_T=40\ex^{(-26354/T)}$ is given by the Arrhenius relation as a function of the temperature $T$ \citep{SawaiEtAl-2013} in units of $cm^{2}\cdotp s^{-1}$, and thus $R_D(T,\bar{r_p})=r_T/(4\pi \bar{r_p}^2)$ is obtained per grain surface area, where $\bar{r_p}$ is the mean grain size.

The energy absorption rate is given by
\begin{equation}
S_D = -\Hcal_D \dot \rho_u = \Hcal_D \frac{A_u}{A_u+2A_w} \dot \rho_p
\label{eq:dehyd7}
\end{equation}

The heat of the reaction $\Hcal_D$ varies slightly with each Serpentine polymorph \citep{WeberGreer-1965}. We adopt a mean value of $4 \cdot 10^9 erg \cdot g^{-1}$.

\subsection{Tidal Dissipation}\label{SS:Tidal}

Tidal deformation and dissipation is a 3-D problem. Since our model is 1-D, we calculate the radial distribution of the dissipated energy based on the global energy dissipation. We make two assumptions in order to approximate the radial distribution of the dissipated heat: (1) the dissipation in the rocky core is small, and (2) the dissipation rate in the icy mantle is proportional to the viscosity of the ice, thus increasing toward the base of the mantle, where the temperature is nearing the melting point of water. The arguments supporting these two assumptions were discussed in section \ref{SSS:TidalHeating}.

The expression for the global tidal dissipated energy per unit volume is given by:
\begin{equation}
S_T = \frac{2 k_2 w^5 R_{tot}^5 e^2}{21 Q V_{tot} G}
\label{eq:tidal_global}
\end{equation}

Where $k_2$ is the dynamic Love number, $Q$ the tidal dissipation factor, $w$ the angular frequency, $R_{tot}$ the total radius, $V_{tot}$ the total volume, $e$ the orbital eccentricity and $G$ the gravitational constant. 

After differentiation of the object, eq. \ref{eq:tidal_global} is used to determine the global tidal energy contribution. We assume (following the analysis of \cite{RobertsNimmo-2008,Roberts-2015}) that a marginal amount of the total energy dissipates in the rocky core, using an arbitrary fudge factor ($f_c$) of order a few percent.

Inside the icy mantle, the total energy is not distributed uniformly, but is assumed to increase radially inward, where the ice is warmer and hence less viscous. The radial distribution is obtained by a weight function $f(r)$ satisfying $\int f(r) dr= 1$. This function is approximated by relying on more detailed studies, for example by \cite{HanShowman-2010}. The rest of the tidal heating parameters are given in table \ref{tab:tidal}.

\begin{table}[h!]
\caption{Tidal Heating Parameters}
\centering
\smallskip
\begin{tabular}{|l|l|l|}
\hline
{\bf Parameter}           			& {\bf Symbol} & {\bf Value} \\ \hline
Angular frequency         			& $w$ & 5.307372$\cdot 10^{-5}~s^{-1}$\\
Orbital eccentricity      			& $e$ & 0.0047\\
Love number/Dissipation factor 	&$k_2/Q$& 0.011 \citep{LaineyEtAl-2012}\\
Gravitational constant           & $G$ & $6.67259 \cdot 10^{-8}~cm^3\cdot g^{-1}~s^{-2}$ \\
\hline
\end{tabular}
\label{tab:tidal}
\end{table}

The spatial parameters related to R and V are not constant, varying throughout the evolution as a result of differentiation, compaction and water ablation.

\subsection{Set of equations}

Considering  the transitions between four different phases of water -- amorphous ice, crystalline ice, liquid and vapor, and between two phases of silicates -- aqueously unaltered and aqueously processed, we have six different components that we denote by subscripts: $u$ - aqueously unaltered rock; $p$ - aqueously processed rock; $a$ - amorphous water ice; $c$ - crystalline water ice; $\ell$ - liquid water; $v$ - water vapor.

The independent variables are: the cumulative volume - $V$; temperature - $T$; densityies - $\rho_a$, $\rho_w=\rho_c+\rho_{\ell}$, $\rho_v$ and $\rho_d$, as well as the mass fluxes $J_v$ (water vapor) and $J_\ell$ (liquid water), as functions of 1-D space and time $t$. The set of equations to be solved is:

\begin{eqnarray}
\label{eq2}
\frac{\di(\rho U)}{\di t} + \frac{\di}{\di V}\left(-K\frac{\di T}{\di V}\right)+\frac{\di(U_vJ_v + U_\ell J_\ell)}{\di V}+q_\ell\Hcal_\ell-S=0 \\
\label{eq3}
\frac{\di \rho_v}{\di t} + \frac{\di(J_v)}{\di V}  = q_v \\
\label{eq5}
\frac{\di \rho_w}{\di t} + \frac{\di J_\ell}{\di V}  = \lambda(T)\rho_a - q_v + \frac{2A_w}{A_u} \left(R_D \rho_p - R_S \rho_u\right) \\
\label{eq7}
\frac{\di \rho_a}{\di t} = -\lambda(T)\rho_a  \\
\label{eq4}
J_v = - \phi_v (1 - \chi) \frac{\di\left(P_v / \sqrt{T}\right)}{\di V}  \\
\label{eq6}
J_\ell = -\phi_\ell \chi \left(\frac{\di(P_\ell)}{\di V}+\rho_\ell g\right) \\
\label{eq1}
Gm\rho = -4\pi (3/4\pi)^{4/3}V^{4/3}\frac{\di P}{\di V}
\end{eqnarray}

Equations (\ref{eq3},\ref{eq5},\ref{eq7}) are the mass conservation equations, where $\lambda(T)$ is the rate of crystallization of amorphous ice, $R_S$ the serpentinization rate, $R_D$ the dehydration rate and $q_v$ the rate of sublimation/evaporation or deposition/condensation respectively. The mass fluxes are given by equations (\ref{eq4}) and (\ref{eq6}), where $\phi_v$ and $\phi_\ell$ are the permeability coefficients. 

In the energy conservation equation (\ref{eq2}), $U$ denotes energy per unit mass, $\Hcal_{\ell}$ the latent heat of fusion (melting), $q_{\ell}$ the rate of melting/freezing, $K$ the effective thermal conductivity, accounting for heat transferred by conduction/convection, while $(U_vJ_v + U_\ell J_\ell)$ accounts for the heat transferred by advection. The sum of all energy sources $S$ includes the energy supplied by crystallization of amorphous ice, the energy lost by sublimation, and all the other possible internal heat sources, such as radiogenic heating, tidal heating, change in gravitational potential energy and heat released or absorbed by geochemical reactions. The last equation, eq. (\ref{eq1}), imposes hydrostatic equilibrium, using the equation of state of \cite{MalamudPrialnik-2015}). All the other variables are easily derived from the independent variables and the volume distribution. The boundary conditions adopted here are straightforward: vanishing fluxes at the center and vanishing pressures at the surface. The surface heat flux is given by the balance between solar irradiation (albedo dependent), thermal emission and heat absorbed in surface sublimation of ice.

The model uses an adaptive-grid technique, specifically tailored for objects that change in mass or volume. Since the body is allowed to grow or shrink (as a result of differentiation, ablation or compaction), a moving, time dependent boundary condition is implemented. The numerical solution is obtained by replacing the non-linear partial differential equations with a fully implicit difference scheme and solving a two-boundary value problem by relaxation in an iterative process. The volume is distributed over a variable $x$ that assumes integer values of $i$, from $i=1$ at the center to $i=I$ at the surface; here we use $I=100$. Thus, an equation is required for determining the grid zoning, to be solved along with the others. Here we use a geometric series for volume distribution over the grid points. Since temporal derivatives are taken at constant $V$, whereas $V=V(x,t)$, the following transformation is implemented in the difference scheme: 

\begin{equation}
\left(\frac{\di}{\di t}\right)_V = \left(\frac{\di}{\di t}\right)_x - \left(\frac{\di V}{\di t}\right)_x \ . 
\left(\frac{\di}{\di V}\right)_t
\end{equation}

\section{Results of Evolutionary Calculations}\label{S:Results}

\subsection{Initial model}\label{SS:initial}

We start with a satellite whose initial homogenous structure is more more massive, more icy and more porous than present-day Enceladus, presuming that it will gradually transform into its present day state as a result of long-term internal heating. We incorporate enough short-lived radionuclides to trigger a warm early evolution. This is equivalent to determining the satellite's formation time. Tidal heating is assumed to be triggered and maintained due to the orbital eccentricity, which can be sustained by the 2:1 mean motion orbital resonance with Dione, although the satellite's motion around the planet is not calculated. The initial composition remains to be determined. Since the inner mid-sized satellites of Saturn have varying fractions of silicates, there is no 'common' rock/ice mass ratio that can be attributed to Enceladus. But given the similar size and orbital proximity of Mimas and Enceladus, it is reasonable to assume, only as a starting working hypothesis, that Mimas and Enceladus had a similar initial composition. 

To obtain the rock/ice ratio for Mimas, we assume that it has remained cold and hence unprocessed during its evolution. This is indicated by its present lack of activity and by its lacking any anomalous crater distribution or any other indication for activity \citep{SchmedemannNeukum-2011}. We thus apply our model with limited heat sources, and find that the observed radius and density of Mimas are best obtained with an initial rock/ice mass ratio of approximately 2. Then we assign the same initial rock/ice mass ratio to Enceladus as well. In Section \ref{S:Discussion} we suggest and test alternative initial compositions.

The initial and physical parameters used in our model are listed in Table \ref{tab:init}. The rock contains the long-lived radionuclides $^{235}U$, $^{40}K$, $^{238}U$ and $^{232}Th$, as well as the short-lived radionuclides $^{26}Al$ and $^{60}Fe$ with initial abundances typical of meteorites. The abundances of short-lived radionuclides depend on the formation time after the formation of CAI. The outcome of early evolution depends not only on radionuclides abundances, but also on the initial temperature, so it is the combination of these two parameters that determines the extent of early 
internal processing. For the initial model we adopt a cold initial temperature of 70$K$ and post CAI formation time of 4.5 $Myr$. A higher initial temperature and later formation time would have identical results.

\begin{table}[h!]
\caption{Initial and physical parameters}
\centering
\smallskip
\begin{tabular}{|l|l|l|}
\hline
{\bf Parameter}                      & {\bf Symbol} & {\bf Value} \\ \hline
Initial uniform temperature          & $T_0$ & 70~K \\
Nominal $^{26}$Al abundance          & $X_0$($^{26}$Al) & $6.7\cdot 10^{-7}$ \\
Nominal $^{60}$Fe abundance          & $X_0$($^{60}$Fe) & $3.46\cdot 10^{-7}$ \\
Nominal $^{235}$U abundance          & $X_0$($^{235}$U) & $6.16\cdot 10^{-9}$ \\
Nominal $^{40}$K abundance           & $X_0$($^{40}$K) & $1.13\cdot 10^{-6}$ \\
Nominal $^{238}$U abundance          & $X_0$($^{238}$U) & $2.18\cdot 10^{-8}$ \\
Nominal $^{232}$Th abundance         & $X_0$($^{232}$Th) & $5.52\cdot 10^{-8}$\\
\hline
Harmonic mean pore size              & $\bar{r_p}$ & 1~cm \\
Saturn's semi-major axis             & $a$ & 9.5820172~AU\\
Ice specific density                 & $\varrho_{a,c}$ & 0.917~g~cm$^{-3}$ \\
Water specific density               & $\varrho_\ell$ &   0.997~g~cm$^{-3}$ \\
Rock specific density (u)            & $\varrho_u$ &  $3.5+2.15\cdot 10^{-12}P$~g~cm$^{-3}$ \\
Rock specific density (p)            & $\varrho_p$ &  $2.9+3.41\cdot 10^{-12}P$~g~cm$^{-3}$ \\
Water thermal conductivity           & $K_\ell$ & $5.5\cdot 10^4$~erg~cm$^{-1}$~s$^{-1}$~K$^{-1}$\\
Ice thermal conductivity (c)         & $K_c$ & $5.67\cdot 10^7/T$~erg~cm$^{-1}$~s$^{-1}$~K$^{-1}$ \\ 
Ice thermal conductivity (a)         & $K_a$& $2.348\cdot 10^2T+2.82\cdot 10^3$\\
                                     &       & erg~cm$^{-1}$~s$^{-1}$~K$^{-1}$\\
Rock thermal conductivity (u)        & $K_u$& $10^5/(0.11+3.18\cdot 10^{-4}T)+$\\
                                     &       &$3.1\cdot 10^{-5}T^3$ erg~cm$^{-1}$~s$^{-1}$~K$^{-1}$\\
Rock thermal conductivity (p)        & $K_p$& $10^5/(0.427+1.1\cdot 10^{-4}T)+$\\
                                     &       &$8.5\cdot 10^{-6}T^3$ erg~cm$^{-1}$~s$^{-1}$~K$^{-1}$\\
\hline
\end{tabular}
\label{tab:init}
\end{table}

\pagebreak[4]

There remains to determine the surface boundary condition, which will define the surface temperature, and consistently with it, the surface heat flux and the vapor flux. This is determined by the heat flux at the surface of the object provided by an external heat source. For an object in orbit around a star, this would be the stellar irradiation. In the case of a satellite, the effect of the planet through tidal heating should be added. Neither of these sources affect the surface uniformly, but for an average rate of heating may be calculated. Here we assume that the boundary conditions have remained constant throughout evolution, and we choose a surface influx that will allow ice loss in an amount that is compatible with the observed rate in the plume of Enceladus (see section \ref{SSS:Plume}), that is, 150 $kg \cdot s^{-1}$. We also choose an initial mass  by adding the mass it lost over its lifetime to its present day mass, which amounts to an increase by about 1/6 its present day mass. This yields a uniform surface temperature of $\sim$135$K$. This choice is largely arbitrary, keeping in mind that the equilibrium temperature at Saturn's distance from the Sun is only $\sim$70$K$ (see, however \cite{TravisSchubert-2015}). It is an attempt to test the possibility of continual mass loss from the surface of an icy satellite. At the distance of Jupiter, for example, this would be the correct temperature to assume. Some justification is provided by the high average surface temperature measured at the SPT, which also led \cite{TravisSchubert-2015} to assume a high temperature boundary condition in their model, albeit restricted to that region.

\subsection{The evolutionary course}\label{SS:Evolution}
The course of evolution is illustrated by a series of figures that represent surface plots: each property is shown as a function of time and of radial distance from the center of the body. Since the radius of the satellite changes during evolution (from about 300 $km$ to about 250 $km$), the upper boundary of the plots changes with time.
 
\begin{figure}[h!]
\begin{center}
\includegraphics[scale=0.54]{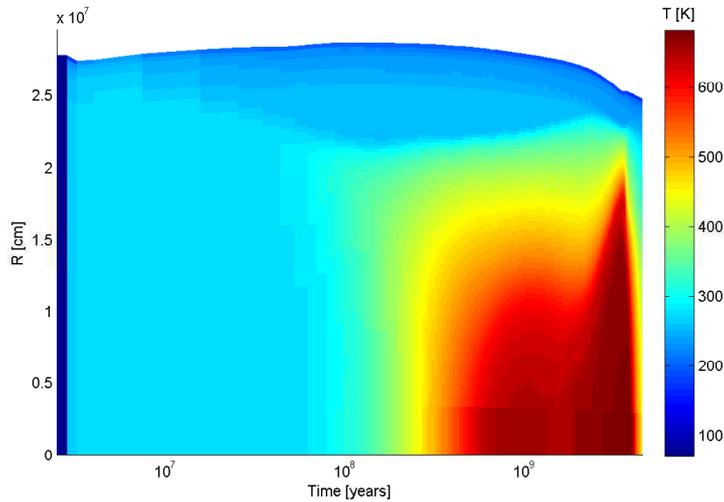}
\caption{Temperature as a function of time and radial distance.}
\label{fig:5T}
\end{center}
\end{figure}

Figure \ref{fig:5T} shows the radial and temporal change in temperature. The evolution starts with a short period of intensive radiogenic heating by short-lived radionuclides, which results in a rapid rise of temperature. As the ice warms, it becomes more susceptible to compaction. As a result, the radius of the body decreases by as much as 20 $km$.

\begin{figure}[h!]
\begin{center}
\subfigure[]{\label{fig:5RODUdivROD}\includegraphics[scale=0.54]{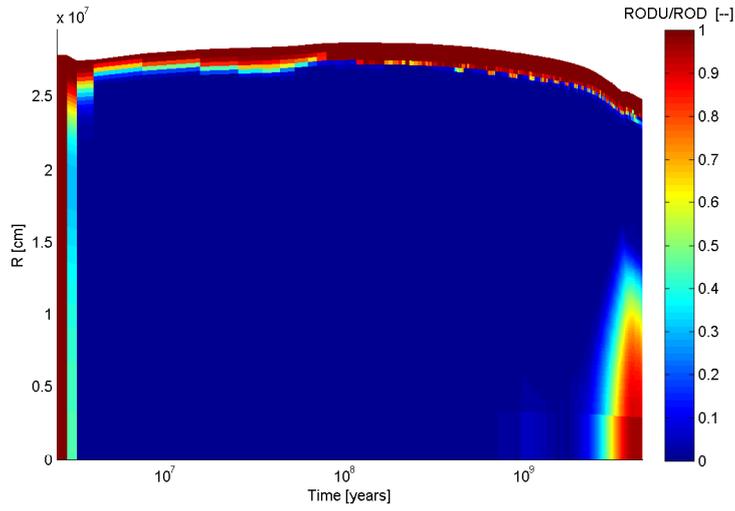}}
\subfigure[]{\label{fig:5ROWdivRO}\includegraphics[scale=0.54]{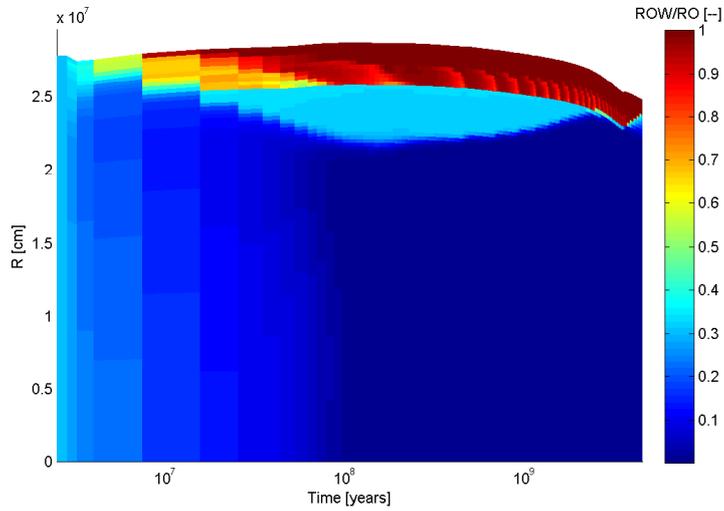}}
\caption{Fraction of aqueously unaltered rock (a) and fraction of  H$_2$O (b) as a function of time and radial distance.}
\end{center}
\end{figure}

Eventually, the temperature in the central part reaches the melting point of ice and liquid water is produced. The water is absorbed in part by the rock in the process of serpentinization. Since the specific density of hydrated rock is lower than that of unprocessed rock, the porosity decreases. The remaining water migrates outward by capillary motion, in the low-gravity environment (in larger bodies, the behavior should be different). Reaching the cold outer region, the water refreezes. In time, the central part becomes completely depleted of ice; the body has become differentiated into a rocky core overlaid by an ice-rich mantle, a few tens of $km$ thick. 

Meanwhile, the temperature continues to rise; although most of the tidal energy is dissipated in the icy mantle, where the viscosity is low, the small fraction that is absorbed in the core is comparable to the radioactive energy released by the long-lived radionuclides. When the temperature in the central part reaches close to 700$K$, the inverse process to serpentinization takes place, the rock exuding the water it had absorbed. The central part of the body becomes dehydrated, so now the stratification is more complex: a compact, dehydrated rock core, a thick shell of hydrated rock overlying it and a dense icy mantle, containing a small fraction of unprocessed rock. This structure is illustrated in Fig.\ref{fig:5RODUdivROD} and Fig.\ref{fig:5ROWdivRO}.

\begin{figure}[h!]
\begin{center}
\includegraphics[scale=0.54]{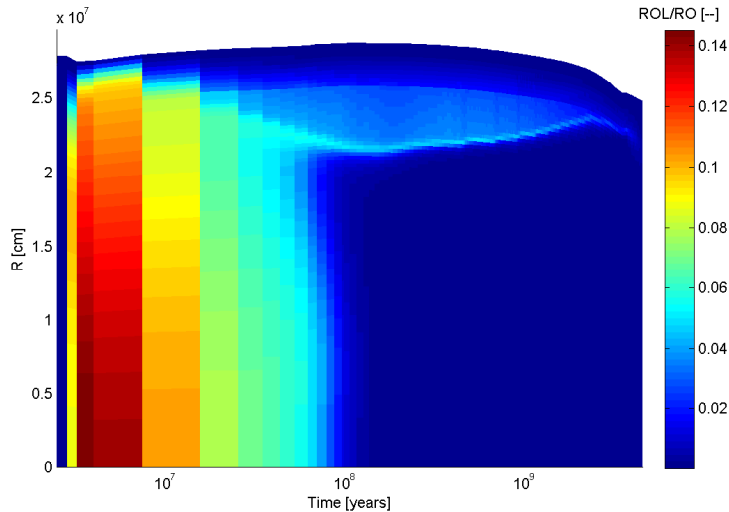}
\caption{Fraction of liquid water as a function of time and radial distance.}
\label{fig:5ROLdivRO}
\end{center}
\end{figure}

The loss of ice at the surface, both by ablation and by outgassing, causes the mantle to thin out slowly. The depletion of ice is countered, but not entirely balanced, by replenishing of water in the mantle due to dehydration reactions in the core. Liquid water, in relatively small amounts (mass fraction), is present at the bottom of the icy outer mantle, as shown in Fig.\ref{fig:5ROLdivRO}.

\begin{figure}[h!]
\begin{center}
\subfigure[]{\label{fig:5RO}\includegraphics[scale=0.54]{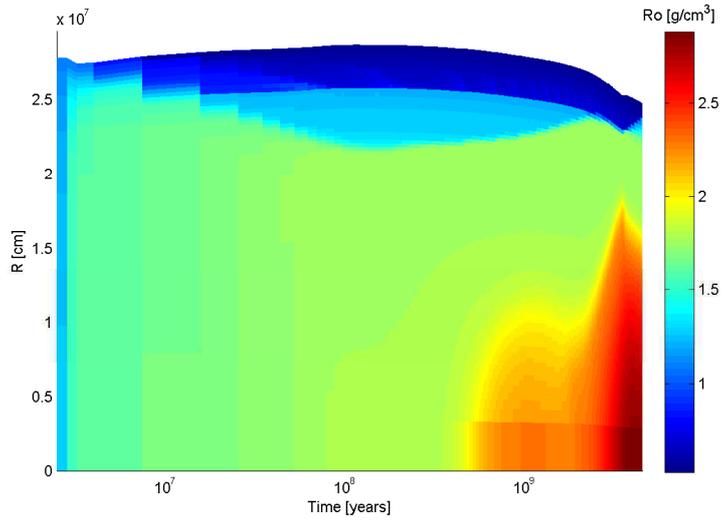}}
\subfigure[]{\label{fig:5PSI}\includegraphics[scale=0.54]{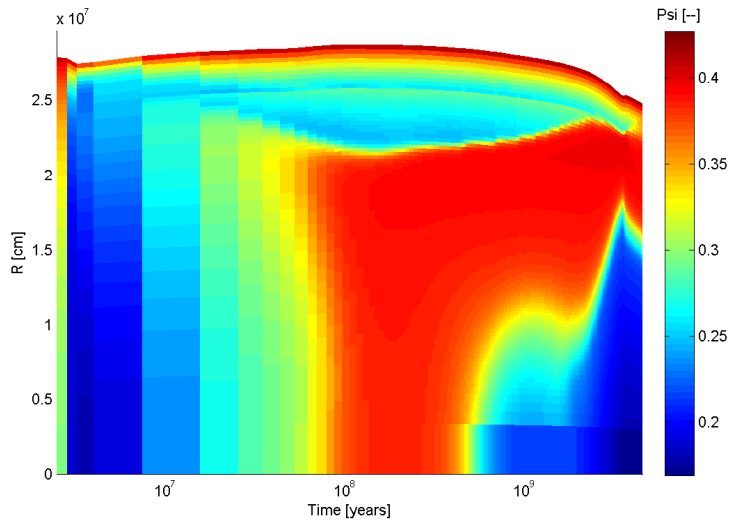}}
\caption{Total density (a) and porosity (b), as a function of time and radial distance.}
\end{center}
\end{figure}

\begin{figure}[h!]
\begin{center}
\includegraphics[scale=0.53]{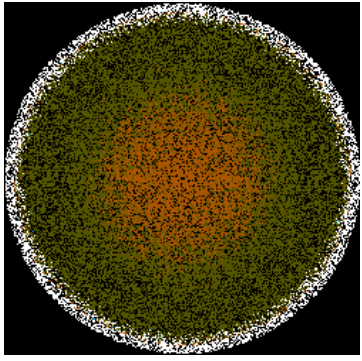}
\caption{Present day cross-section - color interpretation: {\it black} (pores); {\it white} (crystalline ice); {\it brown} (non-hydrated rock); and {\it olive} (hydrated rock)}
\label{fig:5CS}
\end{center}
\end{figure}

We note that at any point in time (vertical line in Fig.\ref{fig:5T}) the temperature decreases monotonically from the center of the body to the surface, and so is the density (see Fig\ref{fig:5RO}). Their profiles, however, vary considerably, as a result of the various processes that take place in the interior of the satellite. The porosity, by contrast, has two peaks, one near the core boundary and another in the outermost layer of the object, as shown in Fig.\ref{fig:5PSI}. At the center, where the temperature and pressure are highest, the porosity is almost vanishing, increasing toward the core boundary. The same trend is exhibited in the icy mantle, which is more compact at the bottom, where the temperature and pressure are higher compared to the surface layers, hence the porosity drops and then increases again closer to the surface.

Overall, thinning of the ice mantle and core compaction result in a gradual decrease of the radius from about 300 $km$ to $\sim250$ $km$ at the end of the evolution run. Finally, the present day structure and composition of the test case for Enceladus, according to the assumptions of the model presented in this section, is shown in Figure \ref{fig:5CS}. The color interpretation is listed in the caption. 

\section{Discussion}\label{S:Discussion}
The free parameters used for the model described in Section \ref{S:Results} were the initial mass, the initial rock/ice mass ratio and the surface temperature (or, equivalently, ice loss rate). Since these were chosen as a test case for Enceladus, we should compare the final model with present-day characteristics derived from observations. These are: the radius of the satellite, or the mean density related to it, and the moment of inertia factor. The estimated values based on observations are 1.61 $g \cdot cm^{-3}$, and 0.335, respectively, while our initial model results yield 0.345 and 1.70 $g \cdot cm^{-3}$, respectively, both values within 6\% of the observed ones. Given the simplifications inherent to a 1-D model, we may regard this as a good agreement.

The free uncertain initial parameters of the model were the formation time (which determines the amount of short-lived radioactive nuclei in the initial state) and the rock/ice mass ratio (which affects the internal structure). In order to test how strongly do the final results depend on the initial choice of these parameters, that is, whether other parameter combinations may lead to successful models, we calculated three additional models (hereafter Models 1-3). 
\begin{itemize}
\setlength{\itemsep}{0pt}
\setlength{\parskip}{0pt}
\item{Model 1 - Initial rock/ice mass ratio: 1.75 ; formation time: 4.50 $Myr$.}
\item{Model 2 - Initial rock/ice mass ratio: 1.50 ; formation time: 4.50 $Myr$.}
\item{Model 3 - Initial rock/ice mass ratio: 1.50 ; formation time: 2.25 $Myr$.}
\end{itemize}

For each model, we tested several values of the most uncertain physical parameters -- the ablation rate and the fraction of tidal energy dissipated in the core -- within acceptable limits, in order to optimize the results. Thus, for Models 2 and 3 we adopted the ablation rate of 200 kg/s (150kg/s for Model 1) and $f_c=0.07$ for Models 1 and 2 and $f_c=0.035$ for Model 3. All other physical and initial parameters were left unchanged.

The evolutionary course of model 1-3 is similar in details to the evolution presented for the initial model in section \ref{S:Results}, the structure of the body at the end of the evolution being differentiated into an icy mantle overlying a rocky core. The final structure of all three models is essentially identical, with a mantle thickness of approximately 30-50 $km$. The results are summarized in Table \ref{tab:summary}. 

\begin{table}[h!]
\caption{Summary of results for all Models}
\centering
\smallskip
\begin{tabular}{|l|l|l|l|}
\hline
{\bf Model}      & {\bf 1}  & {\bf 2} & {\bf 3} \\ \hline
Moment of inertia [--]  & 0.329 & 0.331 & 0.302 \\
Mean density [$g \cdot cm^{-3}$] & 1.59 & 1.60 & 1.63 \\
Ice mantle thickness [$km$] & 30 & 30 & 50 \\
Rocky core (central) density [$g \cdot cm^{-3}$] & 2.96 & 2.74 & 3.48 \\
\hline
\end{tabular}
\label{tab:summary}
\end{table}

Thus relatively small variations in the uncertain initial parameters do not change the picture, which means that it would be difficult to derive either the initial rock/ice mass fraction or the formation time, based on the comparison of model results with observations. On the other hand, the determination of the present structure, such as the outer mantle thickness, appears to be more robust.

One of the encouraging results of the model is that it produces a relatively thin global ice shell. Earlier models of Enceladus (listed in table \ref{tab:mantlethickness}) considered (or computed, in the case of \cite{SchubertEtAl-2007}) a much thicker icy mantle, approximately 80-90 $km$ in size. There are two main reasons for the difference between these models and ours. The first is that other models assumed zero porosity, even if such an assumption seems to contradict experimental compaction results. The second is that other models did not account for mantle thinning due to water ablation.

\begin{table}[h!]
\caption{Icy mantle thickness in other proposed models}
\centering
\smallskip
\begin{tabular}{|l|l|}
\hline {\bf Study}     & {\bf Thickness ($km$)} \\ \hline
\cite{HanEtAl-2012}          & 70-100 \\
\cite{ShojiEtAl-2014,RozelEtAl-2014,TravisSchubert-2015}         & 90 \\
\cite{TobieEtAl-2008,BehounkovaEtAl-2010,BehounkovaEtAl-2012}   & 82 \\
\cite{RobertsNimmo-2008}         & 40-90 \\
\cite{SchubertEtAl-2007}        & 95 \\
\cite{Roberts-2015}        & 46-69 \\
\hline
\end{tabular}
\label{tab:mantlethickness}
\end{table}

Recent gravity field measurements have predicted a total mantle thickness (ice+ocean) around 60 $km$ \citep{IessEtAl-2014}, or even 50 $km$ \citep{McKinnon-2015}. This global ice thickness is incompatible with past evolution models, which is why new ideas are now being considered.

We investigate the possibility of convection in such a thin ice shell. Previous studies show that the ice shell can convect only if it is thicker than $\approx$40 $km$ \citep{BarrMcKinnon-2007,RobertsNimmo-2008}. The precise thickness depends on the rheology and ice grain size (or viscosity, see Figure 3 in \cite{McKinnon-2006} for the range of plausible viscosities). \cite{MitriShowman-2008} place an even higher lower limit on the ice shell thickness at 50 $km$ (when assuming grain sizes of 0.1 $mm$). Our results are compatible with previous studies. Based on a model of parametrized convection, we found that for a thin ice shell, convection is only possible if the effective viscosity is lower than $\approx 10^{14}~Pa \cdot s$. In our models we have chosen a viscosity higher than $10^{14}~Pa \cdot s$ (from the range of  $10^{13}~Pa \cdot s$ to $10^{19}~Pa \cdot s$), leading to conduction alone. \cite{OlginEtAl-2011} also discussed the possibility that the ice shell has temporally evolved in thickness, changing the mode of convection. Testing our models with an ice viscosity of $ 10^{13}~Pa \cdot s$, we have found this to be possible. With such a low ice viscosity, convection was active for almost the entire time, but toward the end of the evolution the icy shell thinned enough to switch to conduction instead, and the final results were not affected by much.

We also tested the effect of tidal energy dissipation in the core. Given that most of the tidal energy is absorbed by the low-viscosity ice-rich mantle, we assumed for the initial model that $5\%$ of the tidal energy is dissipated in the core. If much less than that is assumed, the core temperature does not rise enough for dehydration to occur. If twice as much is assumed, dehydration starts earlier and the amount of dehydrated rock increases. This, of course, affects in turn the density profiles, as well as the amount of water released. In Model 3 we have shown that if a shorter formation time is selected, the early evolution is warmer, and thus the tidal energy dissipated in the core can be greatly reduced, and still produce a viable model. 

Our model also results in a much more silicate rich composition, compared to that suggested by other studies. This is again a consequence of the non-zero porosity assumed here by contrast to the frequently used assumption of zero-porosity. Zero porosity demands a lower rock content for the same bulk density. We already started with a higher rock/ice ratio; the ratio increased during evolution as a result of ice loss.

In conclusion, the long-term evolution calculation of an initially homogeneous satellite, a few hundred km in radius, made of a porous mixture of ice and rock, results in a stratified structure. The core is made of rock, with a high-density, dehydrated inner part and a more porous outer part, made of hydrated rock. The core is overlaid by an ice-rich mantle, a few tens of km thick, at the bottom of which there is a significant fraction of liquid water.

Applied to Enceladus, in terms of mass and bulk density, we find that the mantle is thinner than that obtained by models assuming zero-porosity, and the rock/ice ratio -- higher. This model may serve as basis for developing mechanisms that may explain localized phenomena, such as cryovolcanism.

\newpage


\bibliographystyle{icarus} 
\bibliography{bibfile}     

\end{document}